\documentclass[osajnl,preprint,showpacs]{revtex4}
\DeclareRobustCommand{\baselinestretch{2}}

\begin{document}
\title{Lifetime and hyperfine splitting measurements on the $7s$ and $6p$ levels in Rb}
\author{E. Gomez, S. Aubin, L.A. Orozco*, and G.D. Sprouse}
\affiliation{Dept. of Physics and Astronomy, State University of
New York, Stony Brook, NY 11794-3800 \\ *Present address: Dept. of
Physics, University of Maryland, College Park, MD 20742-4111}

\begin{abstract}
We present lifetime measurements of the $7S_{1/2}$ level and the
$6p$ manifold of Rb.  We use a time-correlated single-photon
counting technique on a sample of $^{85}$Rb atoms confined and
cooled in a magneto-optical trap. The upper state of the
$5P_{1/2}$ repumping transition serves as the resonant
intermediate level for two-photon excitation of the $7s$ level. A
probe laser provides the second step of the excitation, and we
detect the decay of the atomic fluorescence to the $5P_{3/2}$
level at 741 nm. The decay process feeds the $6p$ manifold which
decays to the $5s$ ground state emitting uv photons. We measure
lifetimes of $88.07 \pm 0.40$ ns and $120.7 \pm 1.2$ ns for the
$7S_{1/2}$ level and $6p$ manifold, respectively; while the
hyperfine splitting of the $7S_{1/2}$ level is $282.6 \pm 1.6$
MHz. The agreement with theoretical calculations confirms the
understanding of the wavefunctions involved, and provides
confidence on the possibility of extracting weak interaction
constants from a Parity Non-Conservation measurement.
\end{abstract}
\maketitle

\section{Introduction}
The lifetime of an excited level and its hyperfine splitting are
properties related to the electronic wavefunctions of the atom.
The lifetime, through the matrix elements of allowed transitions,
probes the wavefunctions at large radius, while the hyperfine
splitting samples their value at the nucleus. The comparison of
the two types of measurements with theoretical predictions test
the quality of the computed wavefunctions. The calculation of the
wavefunctions have now reached new levels of sophistication
\cite{johnson03,flambaum04} based on Many Body Perturbation Theory
(MBPT). Those calculations are particularly important in the
interpretation of precision tests of discrete symmetries in atoms:
Parity Non-Conservation (PNC) (see for example the Cs measurements
of Wood {\it et al.} \cite{wood97,wood99}) and Time Reversal
(TR).\cite{lamoreaux}

This paper presents our measurements on $^{85}$Rb atoms in a
magneto-optic trap (MOT) using  time-correlated single-photon
counting techniques. We measure the lifetimes of the
$7s~^{2}S_{1/2}$ level and the $6p$ manifold as well as the
hyperfine splitting of the $7s~^{2}S_{1/2}$ level. The work
complements and aids our program of Fr spectroscopy and weak
interaction physics.\cite{orozco02} We carry out all the Fr
measurements in a trapped and cooled atomic gas. Rb and Fr have
very similar properties and the same trap can be used to capture
either of them by selecting the appropriate
wavelengths.\cite{aubin03a} Having the ability to trap both atoms
helps us understand better our experimental results. The trap is
optimized for Fr and it works on-line with the Superconducting
LINAC at Stony Brook. Our Rb measurements are necessary to fully
understand the systematic effects on our measurements of the
equivalent levels in Fr, $9s, 8p$.\cite{aubin03b,aubin04}

The Rb measurements presented here are an important test of MBPT
calculations in a regime where relativistic effects are not as
important as in heavier atoms such as Fr. Measurements of excited
state atomic lifetimes in the low-lying states of the $s$ and $p$
manifolds enhance our understanding of the wavefunctions and the
importance of correlation corrections in their calculation.

The paper is structured as follows: We present the lifetime
measurements in section \ref{seclifetime}, detail the hyperfine
splitting measurement in section \ref{HF}, and summarize the work
in the context of similar measurements in Rb and Fr in section
\ref{C}.

\section{Lifetime measurements}
\label{seclifetime}

\subsection{Lifetime and matrix elements}
The lifetime of a quantum mechanical system depends on the initial
and final states wavefunctions and the dominant interaction. Since
the electromagnetic interaction in atomic physics is well
understood, radiative lifetimes provide information on atomic
structure.

The lifetime $\tau$ of an excited state is related to the partial
lifetimes $\tau_i$ associated with each of the allowed decay
channels by:

\begin{equation}
1/\tau=\sum_i 1/\tau_i. \label{lifetime}
\end{equation}

The matrix element associated with a partial lifetime between two
states connected with an allowed dipole transition is given
by:\cite{cowan}

\begin{equation}
\frac{1}{\tau_{i}}= \frac{4}{3}\frac{\omega^{3}}{c^{2}}\alpha
\frac{| \langle J \|r\| J' \rangle |^{2}}{2J'+1} ,
\label{matrixelement}
\end{equation}
where $\omega$ is the transition energy divided by $\hbar$, $c$ is
the speed of light, $\alpha$ is the fine-structure constant, $J'$
and $J$ are respectively, the initial and final state angular
momenta, $\tau_i$ is the excited state partial lifetime, and $|
\langle J \|r\| J' \rangle |$ is the reduced matrix element.

The calculation of the radial matrix elements requires the
wavefunctions of the initial and final states involved in the
decay. The contributions of the wavefunction at large distance
become more important due to the presence of the radial operator.
Knowledge of the atomic lifetimes and branching ratios in Rb will
determine the radial matrix elements for the transitions.

\subsection{Sample preparation}
We use a high efficiency magneto-optical trap (MOT) to capture a
sample of Rb atoms at a temperature lower than 300
$\mu$K.\cite{aubin03a} We load the MOT from a Rb vapor produced by
a dispenser in a glass cell coated with a dry film. The MOT
consists of three pairs of retro-reflected beams, each with 15
mW/cm$^2$ intensity, 6 cm diameter (1/e$^2$ intensity) and red
detuned 19 MHz from the atomic resonance. A pair of coil generates
a magnetic field gradient of 6 G/cm. We trap 10$^5$ atoms, with a
diameter of 0.2 mm and a typical lifetime between 5 and 10 s.

Figure \ref{Levelsrb} shows the energy levels of $^{85}$Rb
relevant for the trap, lifetime and hyperfine splitting
measurements. The trapping and cooling are done with a Coherent
899-21 titanium-sapphire (ti-sapph) laser at 780 nm between the
$5S_{1/2}$ $F=3$ and the $5P_{3/2}$ $F=4$ levels. We repump the
atoms that fall out of the cycling transition with a Coherent
899-01 ti-sapph laser at 795 nm between the $5S_{1/2}$ $F=2$ and
the $5P_{1/2}$ $F=3$ levels. A Coherent 899-21 ti-sapph at 728 nm,
the probe laser, completes the two photon transition. We use a
depumper pulse at 780 nm between the $5S_{1/2}$ $F=3$ and the
$5P_{3/2}$ $F=3$ levels before the two photon transition to take
the atoms out of the cycling transition and into the lower
hyperfine ground state.

The atoms in the trap are excited to the $7s$ level using a two
photon transition through the $5P_{1/2}$ level. To increase the
population transfer to the $7s$ level we split the repumper light
into two paths, one going directly to the trap with a large beam
size to optimize the trapping efficiency and combining the other
with the depumper and probe laser focused on the trap to optimize
the excitation. We send 12 mW of probe power, 9 $\mu$W of depumper
power and 2 mW of repumper power focused to a spot size between 1
and 3 mm to increase the excitation intensity.

Figure \ref{Schematic} shows the schematic of the laser system. We
control the power of the lasers going into the trap with
acousto-optic modulators (AOM) and electro-optic modulators (EOM).
We measure the wavelength of the lasers with a wavemeter (Burleigh
WV-1500). We lock the trap laser to $^{85}$Rb using saturation
spectroscopy. We avoid long term frequency drifts on the probe and
repumper lasers by transferring the long term stability of a He-Ne
laser to the two lasers via a computer controlled scanning
Fabry-Perot cavity.\cite{zhao98}

The $7s$ level has four different electric dipole (E1) allowed
decay channels as shown in Fig. \ref{Levelsrb2}. We detect the
direct decay channel from the $7s$ to the $5P_{3/2}$ at 741 nm to
obtain the lifetime of the $7s$ level. The $7s$ level can also
decay to the $6p$ level and from there cascade down to the $5s$
level emitting a photon at 420 (or 422) nm in this last step. We
collect the fluorescence at 420 (or 422) nm that contains
contributions from the lifetime of the $7s$ and $6p$ levels. Using
the results obtained for the lifetime of the $7s$ level we can
extract a lifetime for the $6p$ manifold.

\subsection{Experimental method}

We use the technique of time-correlated single-photon counting to
measure the lifetimes.\cite{oconnor84} This method has been used
in the past to measure lifetimes of atoms in beams,\cite{young94}
vapor cells\cite{hoeling96} and single ions.\cite{devoe94} Our
group has used it to measure the lifetime of the $7p$, $7d$ and
$9s$ levels in Fr\cite{aubin03b,aubin04,simsarian98,grossman00b}
and of the $5p$ levels in Rb.\cite{simsarian98}

The cycle of the measurement has a repetition rate of 100 kHz
controlled with a Berkeley Nucleonics Corporation BNC 8010 pulse
generator and two Stanford Research Systems DG535 pulse delay
generators as shown in Fig. \ref{Timing}. The cycle starts with
0.7 $\mu$s for state preparation. To do this we first take the
atoms out of the cycling transition and into the lower hyperfine
ground level with a depumper beam at 780 nm between the $5S_{1/2}$
$F=3$ and the $5P_{3/2}$ $F=3$ level, (a) in Fig. \ref{Levelsrb}.
Once there, the atoms are excited with the repumper laser to the
$5P_{1/2}$ $F=3$ level, (b) in Fig. \ref{Levelsrb}, and from there
the probe laser takes them to the $7S_{1/2}$ $F=3$ level, (c) in
Fig. \ref{Levelsrb}. We detect fluorescence for 1.6 $\mu$s while
the counting gate is on while keeping all the lasers off during
the last 1.3 $\mu$s. We use the rest of the cycle (8 $\mu$s) for
cooling and trapping of atoms. At the beginning of the cycle we
turn off the trap laser with an acousto-optic modulator
(AOM)(Crystal Technology 3200-144). The trap beam is focused to a
transverse line in the AOM with a cylindrical lens telescope to
avoid damage to the crystal while maintaining a large diffraction
efficiency. This gives a 10:1 on/off ratio for the trap laser in
260 ns. We allow an extra 240 ns to increase the on/off ratio for
the trap laser before we excite the atoms to the $7s$ level for
200 ns. We turn off the probe with two electro-optic modulators
(EOM)(Gs${\rm \ddot{a}}$nger LM0202) and an AOM (Crystal
Technology 3200-144). The two EOM's give a fast turn off for the
probe laser and the AOM improves the long term on/off ratio. The
turn off of the probe laser can be approximated with a half
Gaussian function with a FWHM of 7 ns. We obtain an on/off ratio
of 600:1 after 20 ns. The fast turn off of the EOM's creates
strong radio frequency (RF) emission. The photomultiplier tube
(PMT) amplifiers can pick this emission and create false detection
pulses. We minimize these false events by enclosing the EOMs and
drivers inside a metallic cage in a separate room. Another AOM
(Crystal Technologies 3200-144) turns off the repumper
simultaneously with the probe. We look for fluorescence from the
$7s$ level and the $6p$ manifold for 1.3 $\mu$s. We turn the
trapping beams back on for the rest of the cycle and then repeat
the entire cycle continuously for the duration of the measurement.

A 1:1 imaging system (f/3.9) collects the fluorescence photons
onto a charge coupled device (CCD) camera (Roper Scientific,
MicroMax 1300YHS-DIF). We monitor the trap with the use of an
interference filter at 780 nm in front of the camera. A
beam-splitter in the imaging system sends 50$\%$ of the light onto
a PMT (Hamamatsu R636). An interference filter at 741 nm in front
of the PMT reduces the background light other than fluorescence
from the $7S_{1/2}$ to the $5P_{3/2}$ level. Another independent
1:1 imaging system (f/3.5) monitors the fluorescence from the $6p$
manifold to the $5S_{1/2}$ level with the use of an interference
filter at 420 nm (10 nm bandwidth at 50$\%$) and a PMT (Amperex XP
2020Q).

Figure \ref{Electronics} shows a block diagram of the electronics
used in the detection and processing of $7s$ and $6p$ photon
events. The heart of the electronics is the time-to-amplitude
converter (TAC) (Ortec 467 for the $7s$ photons and Ortec 437 for
the $6p$ photons). The TAC receives an start and an stop pulse and
outputs a voltage proportional to the time separation between
pulses. The pulse generator used to control the timing of the
lasers provides the stop pulse at a fixed delay from the lasers
turn off. A detected fluorescence photon provides the start pulse.
Only one photon can be processed per cycle. The photon pulse
generated by the PMT goes through some processing before reaching
the TAC. It first goes through an Ortec AN106/N amplifier and then
to an Ortec 934 constant fraction discriminator (CFD). The output
of the discriminator is a pulse of fixed shape with a constant
time delay from the input pulse. The pulse then goes through a
linear gate (Ortec LG101/N) which is open only during the
excitation and fluorescence part of the cycle. Starting the TAC
with a fluorescence photon eliminates the cycles with no detected
photons. An histogram of the output of the TAC in a multichannel
analyzer (MCA) (EG$\&$G Trump-8k for the $7s$ photons and Canberra
3502 for the $6p$ photons) displays the exponential decay directly
in real-time.

We calibrate the MCA by replacing the start pulse given by the PMT
with an electronic pulse generated by the pulse generator. We
change the separation between the start and stop pulses in steps
of 100 ns and fit the resulting signal to find both the linearity
and calibration. We verify the uniformity of the MCA channels by
triggering the PMT with random photon events from room light. The
result is a uniformly flat signal consistent with zero slope.

\subsection{$7s$ level analysis}

We measure the lifetime of the $7s$ level through its decay to the
$5P_{3/2}$ level. We keep the number of atoms in the trap low
(about 10$^5$) to reduce density related effects (diameter of the
trap $\sim$ 0.2 mm). We operate with the number of detected
photons per cycle to be much smaller than one. We apply a small
correction to the data to account for preferential counting of
early events.\cite{oconnor84} This correction appears when we have
more than one photon per cycle. The correction, called pulse
pileup correction, is given by

\begin{equation}
{\cal N}'_i=\frac{{\cal N}_i}{1-\frac{1}{n_E}\sum_{j<i}{\cal N}_j}, \label{pileup}
\end{equation}
where ${\cal N}_i$ is the number of counts in channel $i$ of the
MCA, $n_E$ is the total number of cycles, and ${\cal N}'_i$ is the
corrected number of counts for channel $i$. We typically get one
count every 100 cycles (or 1 ms), which corresponds to a
correction smaller than 1$\%$ in the number of counts per channel.

Figure \ref{Decayrbres2} shows the exponential decay obtained for
a 47 minute accumulation along with the fit and residuals. For
times before -200 ns the small signal comes from the trapping
laser light leakage through the interference filter. Between
t=-200 ns and t=0 ns we turn on the excitation beams (probe and
depumper) which gives the fast rise and plateau on the signal
coming both from the fuorescence of the atoms and the leakage from
the two additional lasers, after t=0 ns we turn all the laser
beams off and the only light remaining is the fuorescence from the
atoms. The fit starts 20 ns after the beams turn off and stops
when the signal is equal to the background. The fitting function
is

\begin{equation}
S_{7s}=c_ae^{-t/\tau}+c_b+c_mt, \label{signal7s}
\end{equation}
with $t$ the time (or channel number) and $\tau$ and $c_i$
($i=a,b,m$) the fitting constants. We obtain a background signal
by repeating the experiment without atoms. The lifetime fit is
affected slightly by the presence of a linear background that we
include in the fit. The slope of the background is about 2 counts
per 1000 channels per 1000 seconds of accumulation and comes from
the slow turn off of the trap laser. This particular decay has a
reduced $\chi^2_{\nu}$ of 1.07, where the noise in the number of
counts is statistical ($\sqrt{S}$). A discrete Fourier transform
of the residuals shows no structure.

\subsubsection{Systematic effects}

We search for systematic effects by varying one experimental
parameter at a time and looking for an effect on the obtained
lifetime. Each measurement lasts for about 3000 s. We study first
the effects that the external variables have on the lifetime. Each
measurement is obtained under slightly different conditions. We
fit them independently using the fitting function of Eq.
\ref{signal7s} and make a correlation study between the obtained
lifetime and the external variable for each case.

{\bf Excitation pulse duration}. We change the excitation pulse
duration between 100 and 800 ns. The lifetime is independent of
the initial conditions of the decay. Changing the pulse duration
can modify the initial conditions for the decay.

{\bf Probe intensity}. We vary the probe intensity over a factor
of ten. This is another way in which we can modify the initial
conditions for the decay.

{\bf Magnetic field}. The presence of a magnetic field from the
MOT may influence the measured lifetime mainly through quantum
beats between the Zeeman sublevels. We change the magnetic field
gradient from 4 to 7 Gauss/cm.

{\bf Number of atoms}. We change the number of atoms from
6$\times$10$^4$ to 1$\times$10$^7$. Increasing the number of atoms
will increase the density and produce more collisions between the
atoms as well as permit radiation trapping. These two effects will
modify the lifetime. The photon detection rate also increases with
the atom number. This rate becomes too high for the electronics of
the MCA and the repetition rate for the experiment has to be
reduced to 10 kHz, with a larger pulse pile-up correction.

We quantify the correlation between the measured lifetime and the
external variables by calculating the linear correlation
coefficient. The integral of the probability distribution
associated with the linear correlation coefficient provides the
degree of correlation of the data with the external variable. An
small value for the integral probability means significant
correlation. In all of the above cases the integral probability of
the linear correlation coefficient is larger than 5$\%$,
consistent with no correlation. We keep the number of atoms low
for all the measurements to avoid systematic effects related to
collisions or to pulse pile-up. Radiation trapping can be ignored
due to the small population in the $5P_{3/2}$ level.

Other effects can influence the measurement but they are not
related to a simple variable as above. In this case we make a
reference measurement and then we modify something to test each
one of the above potential effects to obtain another measurement.
We fit each independent data file using Eq. \ref{signal7s} and
perform an $\chi^2$ test to the obtained lifetimes to find out if
they are consistent with statistical fluctuations.

{\bf Repumper turn off}. We look for an effect of an imperfect
turn off of the repumper light by leaving the repumper on
continuously. The repumper is used as the first step of the two
photon transition and when combined with an imperfect turn-off of
the probe laser it can introduce a false signal.

{\bf Hyperfine level}. To our accuracy level, the lifetime should
be independent of the hyperfine level. We change the initial
hyperfine level of the decay, that is, instead of preparing the
atoms in the $7S_{1/2}$ $F=3$ state we prepare them in the
$7S_{1/2}$ $F=2$ state.

{\bf Electronics}. We look for effects related to the electronic
components by interchanging the MCA for the $7s$ and $6p$
detection systems. It is important to keep the Canberra MCA count
rate low.

{\bf Probe turn off}. An imperfect extinction of the probe laser
will show up as an excess in the initial data points of the decay.
This effect can be revealed by changing the initial/final point
for the fit. The spread of the lifetimes as a function of the
starting point of the fit is consistent with the statistical
uncertainty. There is no dependence on the final point for the fit
within our statistical precision.

All the above measurements give an integral probability for the
$\chi^2$ between 5$\%$ and 95$\%$, consistent with statistical
fluctuations.

{\bf Trap displacement}. We displace the trap keeping the magnetic
field fixed, such that the atoms are sampling a different magnetic
environment. To move the trap position we insert a piece of glass
in front of one of the retro-reflection mirrors in the MOT. We
repeat the same procedure for the three retro-reflection mirrors
in the MOT. The MOT image on the camera shows trap displacements
smaller than one trap diameter in the transversal direction to the
camera and we have no information on the longitudinal
displacement. This is a complex systematic effect since it
involves the change of several experimental parameters such as the
alignment of the excitation lasers. This makes it difficult to
assign a single parameter responsible for the variations we
observe. We tried different combinations of moving the trap with
and without realignment. The integral probability of the $\chi^2$
shows fluctuations larger than statistical. We include an
uncertainty contribution of $\pm$0.38$\%$, equal to the dispersion
of the lifetime values (Fig. \ref{Move}).

{\bf Quantum beats}. We look for quantum beats in the residuals of
the fit. A discrete Fourier transform of the residuals shows no
structure. The value of $\pm$0.1$\%$ quoted for the uncertainty
due to quantum beats comes from a theoretical calculation with a
simple model which assumes well defined Zeeman sublevels as in the
presence of a uniform magnetic field. The presence of a magnetic
field gradient further reduces the quantum beat contribution.

Some of the information obtained can be extended to measurements
in Fr. Changing the number of atoms is complicated in Fr so we can
use the results for Rb to know if we are working in a good regime.
Both atoms have similar atomic structure, so most tests should
give similar results. The most important difference is their
sensitivity to magnetic effects because of the difference in
multiplicity of Zeeman sublevels and hyperfine separation.

\subsubsection{Result and comparison with theory}

The average of the reduced $\chi_{\nu}^2$ of the individual files
is 1.04$\pm$0.08. The different lifetimes from the fit are
averaged to obtain the final result. The lifetime of the $7s$
level is 88.07$\pm$0.38 ns. A fit to the file resulting from
adding all the files gives consistent results. Table
\ref{errorbudget} summarizes the error budget for the experiment.

Figure \ref{Theory} is a comparison of our result with theoretical
predictions\cite{safronova03,theodosiou84} for the lifetime of the
$7s$ level, as well as previous
measurements.\cite{marek80,bulos76} Theoretical calculations are
reaching a level of precision below 1$\%$.\cite{derevianko01} An
experimental verification of this precision is important to
increase the confidence in such calculations. This information is
crucial to extract weak interaction physics out of parity
non-conservation experiments. The prediction from {\it ab initio}
calculations for the $7s$ level lifetime is 88.3
ns.\cite{safronova03} The agreement shows the remarkable level of
sophistication of atomic structure calculations.

\subsection{$6p$ level analysis}

The $7s$ level has the four decay channels shown in Fig.
\ref{Levelsrb2}. We detect the indirect decay from the $6p$ to the
$5s$ levels to obtain information about the $6p$ level. The atoms
decaying from the $6p$ level come from a cascade decay from the
$7s$ level and the decay cannot be described with a single
exponential. The signal from this indirect decay is the sum of
three exponential functions with lifetimes corresponding to the
$7S_{1/2}$, $6P_{1/2}$ and $6P_{3/2}$ levels. We can make use of
the result obtained in the previous section for the lifetime of
the $7S_{1/2}$ level to measure the lifetime of the $6p$ manifold.
Here we present the analysis of the $6p$ signal that contains
contributions from the two fine levels. The fine separation of the
$6p$ levels in Rb is 1.4 nm which is smaller than the 10 nm
transmission width of the interference filter. In the case of Fr,
the fine separation of the corresponding levels is larger and we
use interference filters to resolve both contributions
seperately.\cite{aubin03b,aubin04}

The lifetimes of the two fine $6p$ levels are expected to be
similar. We assume that the decay signal is given by the sum of
two exponential functions, one for the $7s$ level and the other
for the $6p$ level, so that the fitting function is

\begin{equation}
S_{6p}=A_b+A_{7s}e^{-t/\tau_{7s}}+A_{6p}e^{-t/\tau_{6p}}, \label{signal}
\end{equation}
where $\tau_{7s}$ is the lifetime of the $7s$ level obtained in
the previous section, and  $\tau_{6p}, A_b, A_{7s}$ and $A_{6p}$
are the fitting constants. Fig. \ref{Decayresrb6p} shows the
signal obtained for a single file and the resulting curve if we
subtract the background and the exponential contribution from the
$7s$ level. This last curve corresponds to the exponential decay
of the $6p$ manifold. We only use files with low count rates to
avoid systematic effects associated with the slow response of the
MCA.

The lifetime we obtain for the $6p$ manifold depends on the value
of the lifetime of the $7s$ level. The uncertainty in the $7s$
lifetime influences the precision with which we can extract the
$6p$ lifetime. The probability distribution for $\tau_{6p}$ is
given by

\begin{equation}
P(\tau'')= \int d\tau' \frac{1}{\sqrt{2\pi} \sigma_{7s}}
e^{-\frac{1}{2} \left(\frac{\tau'-\tau_{7s}}{\sigma_{7s}}
\right)^2} \frac{1}{\sqrt{2\pi} \sigma_{6p}(\tau')}
e^{-\frac{1}{2}
\left(\frac{\tau''-\tau_{6p}(\tau')}{\sigma_{6p}(\tau')}
\right)^2}. \label{ptau6p}
\end{equation}
The integrand contains two Gaussian distributions, the first one
gives the probability distribution for $7s$ level lifetime
centered on $\tau_{7s}$ with an uncertainty $\sigma_{7s}$ and the
second one gives the probability distribution for the $6p$
manifold lifetime centered on $\tau_{6p}(\tau')$ with an
uncertainty $\sigma_{6p}(\tau')$. We assume a value $\tau'$ for
the lifetime of $\tau_{7s}$ and include that in the fitting
function (Eq. \ref{signal}) to obtain a value for
$\tau_{6p}(\tau')$. We repeat the same procedure for different
values of $\tau'$ and perform the integral of Eq. \ref{ptau6p}.

The result for the integral when $\tau_{6p}$ and $\sigma_{6p}$ do
not strongly depend on $\tau'$ gives approximately
$\tau_{6p}=\tau_{6p}(\tau_{7s})=120.7$ ns and
$\sigma_{6p}=\sqrt{\sigma_{6p}(\tau_{7s})^2+
\left(\frac{d\tau_{6p}(\tau')}{d\tau'} \sigma_{7s}\right)^2}=0.35$
ns, as confirmed by numerical integration. This result assumes
uncorrelated errors for the individual files used on the fit but
includes the spread brought by systematic checks on the $7s$ level
lifetime. The uncertainty in the MCA calibration is at the
0.94$\%$ level. Table \ref{errorbudget6p} summarizes the error
budget that gives a final result for the lifetime of the $6p$
manifold of 120.7$\pm$1.2 ns.

\subsubsection{Simple model}

We can make a comparison between the predicted and the measured signal to give some
bounds on the possible values for the lifetime of each fine level.

The decay signal is obtained by solving the following rate equations

\begin{eqnarray}
& & \frac{dN_{s}}{dt}=-\frac{N_{s}}{\tau_{s}}, \nonumber \\ & &
\frac{dN_{p1}}{dt}=B_{p1}\frac{N_{s}}{\tau_{s}} -\frac{N_{p1}}{\tau_{p1}}, \nonumber
\\ & & \frac{dN_{p3}}{dt}=B_{p3}\frac{N_{s}}{\tau_{s}}
-\frac{N_{p3}}{\tau_{p3}}, \label{evolution}
\end{eqnarray}
where $N_i$ and $\tau_i$ give the population and lifetime
respectively of level $i$, with $i=s,p1,p3$ representing the
$7S_{1/2}$, $6P_{1/2}$ and $6P_{3/2}$ levels, and $B_{p1}=0.132,
B_{p3}=0.255$ the theoretical branching ratios from the $7S_{1/2}$
level to the $6P_{1/2}$ and $6P_{3/2}$
respectively.\cite{safronova03} To solve this equations we need
the initial conditions for the level populations at the beginning
of the decay (or equivalently at the end of the excitation pulse).
Fig. \ref{Decayrbres2} shows that during the excitation the
population of the $7s$ level reaches an steady state very fast. We
will assume the $7s$ population to be constant during the
excitation pulse. We also assume that before the excitation we
have no population in the $6p$ level. With these assumptions we
can calculate the population of the $6p$ levels during the
excitation given by

\begin{eqnarray}
N_{p1}=N_{s}B_{p1}\frac{\tau_{p1}}{\tau_{s}} (1-e^{-t/\tau_{p1}}),\nonumber
\\ N_{p3}=N_{s}B_{p3}\frac{\tau_{p3}}{\tau_{s}}
(1-e^{-t/\tau_{p3}}). \label{initial}
\end{eqnarray}
The excitation lasts for $T$=200 ns, so evaluating these
expressions after this time will give the initial conditions for
the decay. Solving Eq. \ref{evolution} with these initial
conditions gives the population of the three levels as a function
of time. The signal measured by the PMT is proportional to the sum
of the decay rates of each of the $6p$ levels to the $5s$ level.
The underlying assumption that the response of the PMT and the
interference filter is the same for both of the $6p$ levels is
reasonable due to the small energy separation between them (1.4
nm). The signal ($\widetilde{S_{6p}}$) coming out of the PMT is
given by

\begin{eqnarray}
\widetilde{S_{6p}} & &
=\widetilde{A_b}+\widetilde{A}\left(b_{p1}\frac{N_{p1}}{\tau_{p1}}
+b_{p3}\frac{N_{p3}}{\tau_{p3}}\right), \nonumber
\\ & & =\widetilde{A_b}+\widetilde{A'}\left[\left(\frac{b_{p1}B_{p1}}{\tau_{s}-\tau_{p1}}
+\frac{b_{p3}B_{p3}}{\tau_{s}-\tau_{p3}}\right)e^{-t/\tau_{s}}
+b_{p1}B_{p1}\left(\frac{1-e^{-T/\tau_{p1}}}{\tau_{s}}
-\frac{1}{\tau_{s}-\tau_{p1}}\right) e^{-t/\tau_{p1}}\right. \nonumber
\\ & & \left. +b_{p3}B_{p3}\left(\frac{1-e^{-T/\tau_{p3}}}{\tau_{s}}
-\frac{1}{\tau_{s}-\tau_{p3}}\right) e^{-t/\tau_{p3}}\right], \label{signal2}
\end{eqnarray}
where $b_{p1}=0.194, b_{p3}=0.236$ are the branching ratios for
the decays from the $6P_{1/2}$ and $6P_{3/2}$ to the $5s$ level
respectively\cite{safronova03} and $\widetilde{A_b},
\widetilde{A}$ and $\widetilde{A'}$ the background and scale
constants.

To compare this expression with Eq. \ref{signal} we need to
combine the two exponential functions for the $6p$ levels into a
single one since we do not have enough resolution to separate
them, that is we need to make

\begin{equation}
b_{p1}B_{p1}\left(\frac{1-e^{-T/\tau_{p1}}}{\tau_{s}}
-\frac{1}{\tau_{s}-\tau_{p1}}\right) e^{-t/\tau_{p1}}
+b_{p3}B_{p3}\left(\frac{1-e^{-T/\tau_{p3}}}{\tau_{s}}
-\frac{1}{\tau_{s}-\tau_{p3}}\right) e^{-t/\tau_{p3}} \sim C e^{-t/\tau'_{6p}}.
\label{expred}
\end{equation}

The two expressions above will be equal in the least squares
sense, meaning that we will solve for the values of C and
$\tau'_{6p}$ that minimize the square of the difference of the two
sides of the equation in the range from 0 to $\infty$. The
theoretical values for the $6p$ fine level lifetimes are
$\tau_{p1}=129$ ns, $\tau_{p3}=118$ ns.\cite{safronova03} Using
these values we get the following expression for the signal

\begin{equation}
\widetilde{S_{6p}}=\widetilde{A_b}+\widetilde{C}(e^{-t/\tau_{s}}-1.29e^{-t/\tau'_{6p}}),
\label{signal3}
\end{equation}
with $\tau'_{6p}=120.7ns$. The ratio of the amplitudes of the $7s$
and $6p$ exponential functions is fixed by this model. Using the
fitting parameters from Eq. \ref{signal} for the experimental
result we obtain $A_{6p}/A_{7s}=-1.44\pm0.01$. The difference
between the predicted ratio and the one obtained is 12$\%$.

We can invert the previous procedure to set limits on the possible
values of the $6p$ fine level lifetimes. If we take $\tau'_{6p}$
equal to the experimental value (or some other value) we can only
obtain that value with specific combinations of $\tau_{p1}$ and
$\tau_{p3}$. This will not fix either $\tau_{p1}$ or $\tau_{p3}$,
but it will create a functional relation between the two. Fig.
\ref{6pprediction} gives the 1$\sigma$ and 2$\sigma$ bands for the
experimental result using the described method with the branching
ratios assumed to be constant. The theoretical predictions are
also included in the figure and the {\it ab initio}
calculation\cite{safronova03} is in agreement with the
experimental result that includes the statistical and calibration
uncertainty.

\section{Hyperfine splitting}
\label{HF}

\subsection{Hyperfine splitting and matrix elements}
The hyperfine splitting in an atom is produced by the interaction
of the electrons with the nuclear magnetic moment. The hyperfine
splitting constant for an $s$ state is given by \cite{kopfermann}

\begin{equation}
A=\frac{8\pi}{3} \frac{\mu_0\mu_B}{4\pi h} 2g\mu_N |\psi(0)|^2
\kappa, \label{acte}
\end{equation}
where $\mu_0$ is the magnetic constant, $\mu_B$ is the Bohr
magneton, $\mu_N$ is the nuclear magneton, $g$ is the nuclear
g-factor and $\kappa$ is a correction term that includes the
relativistic correction, the Breit correction and the
Bohr-Weisskopf effect.

The hyperfine splitting constant works as a probe for the magnetic
environment created by the electrons at the nucleus. Measurements
of the hyperfine splitting will tests the wavefunctions at short
distances.

The experimental setup used for the lifetime measurements gives
the flexibility to reach both of the $7s$ hyperfine levels. We
have a clean detection method for the number of atoms promoted to
the $7s$ level through the fluorescence photons from the $7s$
level or from the $6p$ manifold. In this section we present the
results for the measurement of the hyperfine splitting of the $7s$
level.

\subsection{Experimental method}

We measure the hyperfine splitting of the $7s$ state by scanning
the frequency of the probe laser and counting the number of
photons as a function of frequency. The excitation sequence
corresponds to the one used for the lifetime measurement with the
excitation pulse length increased to 1.5 $\mu$s. We monitor the
wavelength of the probe laser with a wavemeter (Burleigh WV-1500)
that has a resolution of $\pm$30 MHz. We improve the meaurement
resolution with a Fabry-Perot cavity which acts as a frequency
ruler. We send the probe laser and a frequency stabilized
Melles-Griot He-Ne laser (05-STP-901) into a Fabry-Perot cavity
that is constantly scanning. We detect and digitize the
transmitted intensity. A computer monitors the position of the
transmission peak of the probe laser relative to two transmission
peaks of the He-Ne laser. Using this method we control the drift
of our lasers to less than 1 MHz per hour.\cite{zhao98}

As we scan the probe laser, its relative position with respect to
the He-Ne peaks will change and may even move to a neighboring
free spectral range. Knowledge of the free spectral range of the
cavity gives us a ruler to measure frequency differences.

We calibrate the cavity with an EOM (New Focus 4002) driven with a
signal generator (Giga-tronics 1026) to add sidebands of known
frequency to the probe laser before it enters the cavity. We
select the probe laser frequency equal to one of the hyperfine
levels and the frequency driving the EOM about half of the
hyperfine splitting, such that the second order sideband is close
to the other hyperfine level. A scan of the sideband frequency
around this value gives a local cavity calibration to $\pm$0.42
MHz.

The method for detection of fluorescence photons is the same as
the one used for the lifetime measurement (Fig. \ref{Electronics})
with the TAC and MCA replaced by a gate and delay generator (Ortec
416A) to create positive pulses and a multichannel scaler (MCS)
(National Instruments BNC 2090) to count the number of detected
photons per second.

\subsection{Analysis and results}

The resolution of the wavemeter can be improved if one assumes
that the noise is Gaussian. Fig. \ref{Wavemeter} shows a plot of
the number of photons vs wavemeter reading. The origin is
arbitrarily defined to be 13732.476 cm$^{-1}$ on the wavemeter. We
fit the data with two Gaussian peaks plus a background. With this
method we find a hyperfine separation of 277.3$\pm$5.4 MHz.

We perform several scans recording both the number of counts and
the relative (or percent) position of the laser transmission peak
with respect to two fixed He-Ne transmission peaks on the cavity.
The result of a typical scan is shown in Fig. \ref{Cavityscan}.
The two peaks correspond to the two hyperfine levels and they are
separated by one free spectral range. We fit each peak with a
Lorentzian function plus a background and then average over all
the scans. The difference in position between the two peaks is
compared against the calibration to obtain the separation in MHz.
The statistical uncertainty is the main contribution on the error
budget (Table \ref{errorbudgethfs}) with a 0.46$\%$ contribution.

The presence of a magnetic field may modify the hyperfine
splitting measurement through a Zeeman splitting of the magnetic
sublevels. Assuming all the atoms start from a common state and
reach the highest magnetic sublevel on each of the hyperfine
levels we obtain an upper limit for the contribution of the Zeeman
shift of 0.16$\%$.

The presence of laser beams on the excitation can induce an AC
shift and splitting of the hyperfine levels. We do not observe any
clear asymmetry or splitting on each of the hyperfine peaks,
although we do see some power broadening. The natural linewidth
from the lifetime is 11.4 MHz, whereas the data has a linewidth of
24 MHz which shows power broadening. We model the scan signal by
solving the steady state optical Bloch equations\cite{grossman00a}
and obtain an spectrum consistent with the data (Fig.
\ref{Simulation}). The intensities and detunings of the beams were
adjusted to approximate the data and are consistent with the
experimental ones. Using this model, we set limits for the effect
of the AC Stark shift on the hyperfine splitting we measure to
less than 0.2$\%$.

Table \ref{errorbudgethfs} summarizes the error budget for the
measurement. We find a hyperfine splitting for the $7s$ level of
282.6$\pm$1.6 MHz.

The relation between the hyperfine shift and the magnetic dipole
hyperfine constant (A) for an s level is given by

\begin{equation}
\frac{E_{HF}}{h}=A\frac{K}{2}, \label{aconstant}
\end{equation}
with $K=F(F+1)-I(I+1)-J(J+1)$. In our case $I$=5/2 and $J$=1/2 so
we have $A$=94.2$\pm$0.6 MHz. Fig. \ref{Theoryhfs} shows a
comparison of the present work with previous
experiments\cite{gupta73} and a theoretical
prediction.\cite{safronova99} The theoretical prediction assumes a
nuclear magnetic moment of 1.3534 in units of $\mu_N$ the nuclear
magneton. We find good agreement between both experimental results
and theory. Measurements of the hyperfine splitting of an $s$
level are useful to understand the contributions from radiative
corrections such as the one produced by the Breit
interaction.\cite{sushkov01}

\section{Conclusions}
\label{C} We have measured the lifetime and hyperfine splitting of
the second excited $S_{1/2}$ level of Rb and the lifetime of the
second excited $p$ manifold of Rb. We have used two-photon
excitation and time-correlated single-photon counting techniques
on a sample of cold $^{85}$Rb atoms confined in a MOT. Our $7s$
lifetime measurement has excellent statistics and the result is
limited by systematic uncertainties. The measurement represents a
ten-fold improvement in the accuracy from previous measurements.
The lifetime tests calculations of radial matrix elements
connecting excited states in Rb. Comparisons with {\it ab-initio}
calculations of the matrix elements for the different decay
channels agree to better than 0.3\%. The measurement of the
lifetime of the $6p$ manifold does not differenciate between the
two decay channels from the fine structure and achieves less
accuracy, while a comparison to theory is model dependent, but
sets bounds for the two contributions. The $0.57\%$ hyperfine
splitting measurement is in agreement with previous values and
theoretical calculations.

All these measurements confirm the high quality predictions of
MBPT calculations and increase the confidence in the methods
applied to heavier alkali atoms such as Fr and Cs for similar
spectroscopic studies and extraction of weak interaction
information from PNC measurements.

\section*{Acknowledgments}

Work supported by NSF. E. G. acknowledges support from CONACYT. We
thank M.S. Safronova for preliminary unpublished results.

\newpage
\begin{table}[h]
{\bf \caption{\label{errorbudget}Error budget for $7s$ level
lifetime measurement.}}
\begin{center}
\begin{tabular}{lr}
\hline
 Error & $[\%]$ \\ \hline
 Statistical & $\pm 0.17$ \\
 Trap displacement & $\pm 0.38$ \\
 Time calibration & $\pm0.01$ \\
 TAC/MCA nonlinearity & $\pm0.04$ \\
 Quantum beats & $<\pm 0.10$ \\
\hline
 {\bf Total} & {\bf $\pm$0.43} \\
\hline
\end{tabular}
\end{center}
\end{table}

\newpage

\begin{table}[h]
{\bf \caption{\label{errorbudget6p}Error budget for $6p$ manifold
lifetime measurement.}} \begin{center}
\begin{tabular}{lr}
\hline
Error & $[\%]$ \\ \hline
 Statistical & $\pm 0.15$ \\
 $7s$ uncertainty propagation & $\pm 0.25$ \\
 Time calibration & $\pm0.94$ \\
\hline
 {\bf Total} & {\bf $\pm$0.98} \\
\hline
\end{tabular}
\end{center}
\end{table}

\newpage

\begin{table}[h]
{\bf \caption{\label{errorbudgethfs}Error budget for $7s$ level
hyperfine splitting measurement.}}
\begin{center}
\begin{tabular}{lr}
\hline
Error & $[\%]$ \\ \hline
 Statistical & $\pm 0.46$ \\
 Cavity calibration & $\pm 0.15$ \\
 Differential Zeeman shift & $\pm 0.16$ \\
 AC Stark asymmetrical broadening & $<\pm 0.20$ \\
\hline
 {\bf Total} & {\bf $\pm$0.55} \\
\hline
\end{tabular}
\end{center}
\end{table}

\newpage

\section*{List of Figure Captions}

\begin{figure}[h]\centerline{\scalebox{1}{\includegraphics{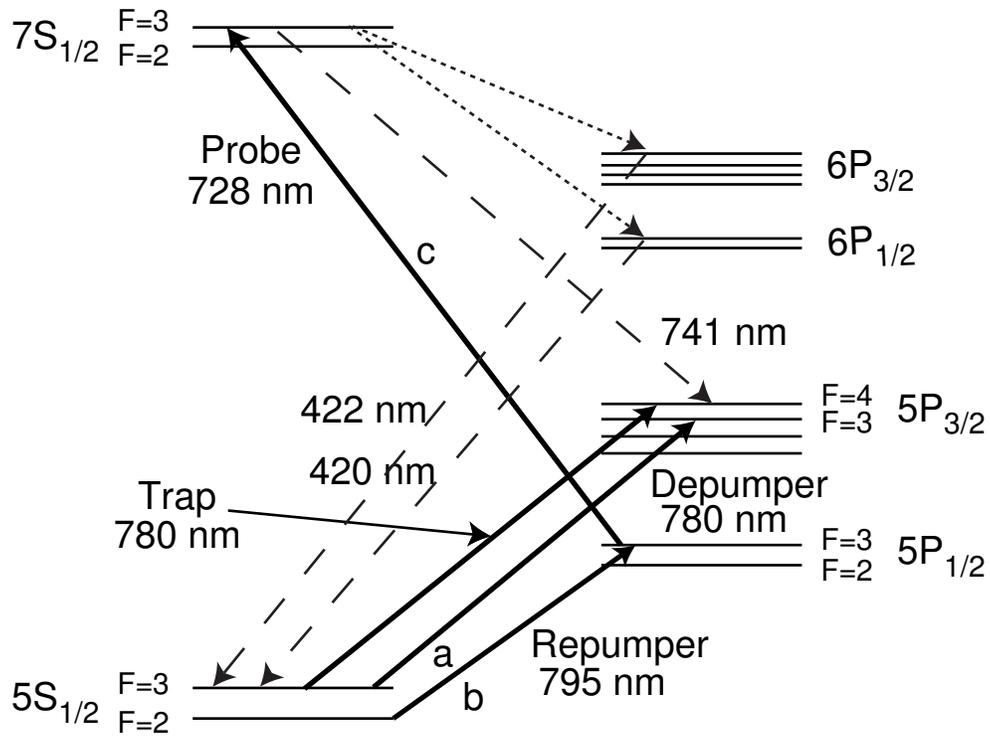}}}
\caption{Energy levels of $^{85}$Rb for trapping and two photon
excitation to the $7s$ level (solid lines) fluorescence detection
(dashed line) and undetected fluorescence (dotted line).
gomezF1.EPS. \label{Levelsrb}}
\end{figure}

Fig. \ref{Levelsrb}. Energy levels of $^{85}$Rb for trapping and
two photon excitation to the $7s$ level (solid lines) fluorescence
detection (dashed line) and undetected fluorescence (dotted line).

\begin{figure}[h]\centerline{\scalebox{1}{\includegraphics{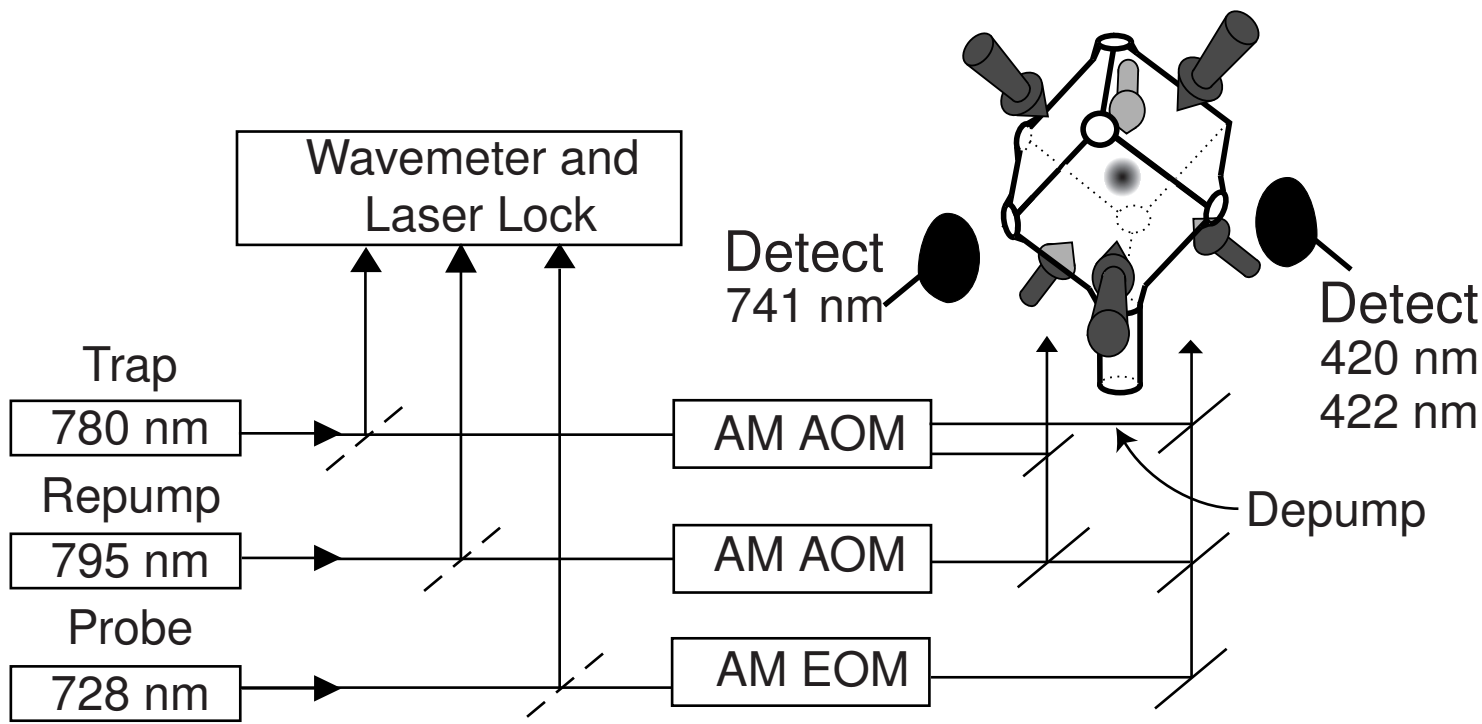}}}
\caption{Schematic of the trap. AM EOM stands for amplitude
modulation with an electro-optic modulator and AM AOM for
amplitude modulation with an acousto-optic modulator. gomezF2.EPS.
\label{Schematic}}
\end{figure}

\noindent Fig. \ref{Schematic}. Schematic of the trap. AM EOM
stands for amplitude modulation with an electro-optic modulator
and AM AOM for amplitude modulation with an acousto-optic
modulator.

\begin{figure}[h]\centerline{\scalebox{1}{\includegraphics{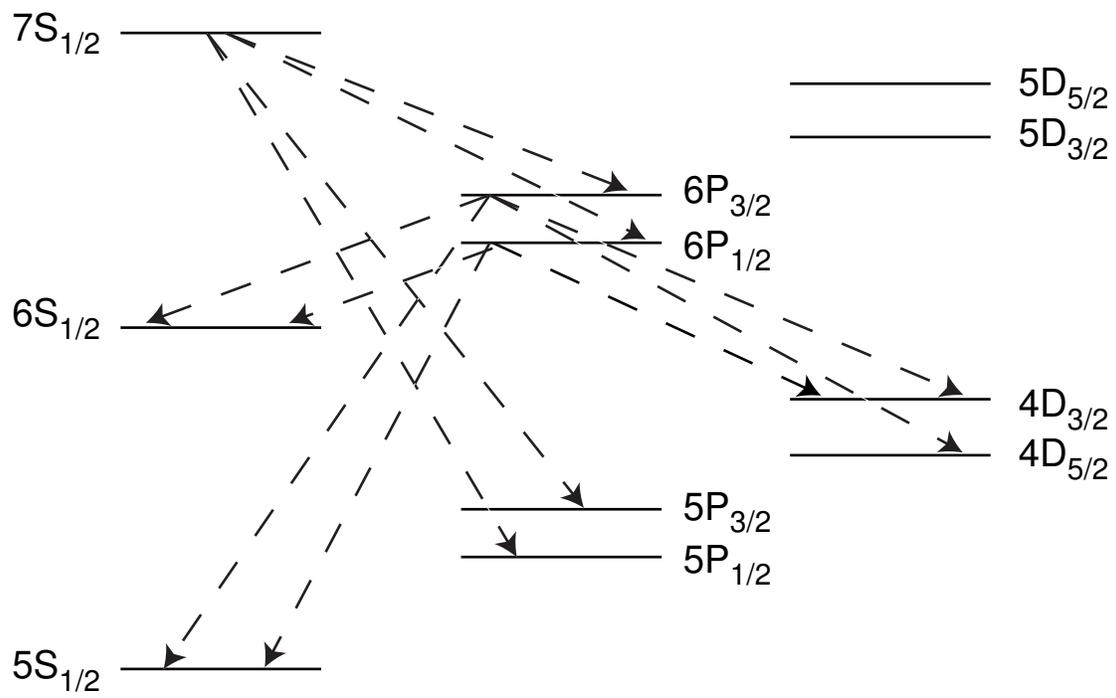}}}
\caption{Decay paths for the $7s$ and $6p$ levels of $^{85}$Rb.
gomezF3.EPS. \label{Levelsrb2}}
\end{figure}

\noindent Fig. \ref{Levelsrb2}. Decay paths for the $7s$ and $6p$
levels of $^{85}$Rb.

\begin{figure}[h]\centerline{\scalebox{1}{\includegraphics{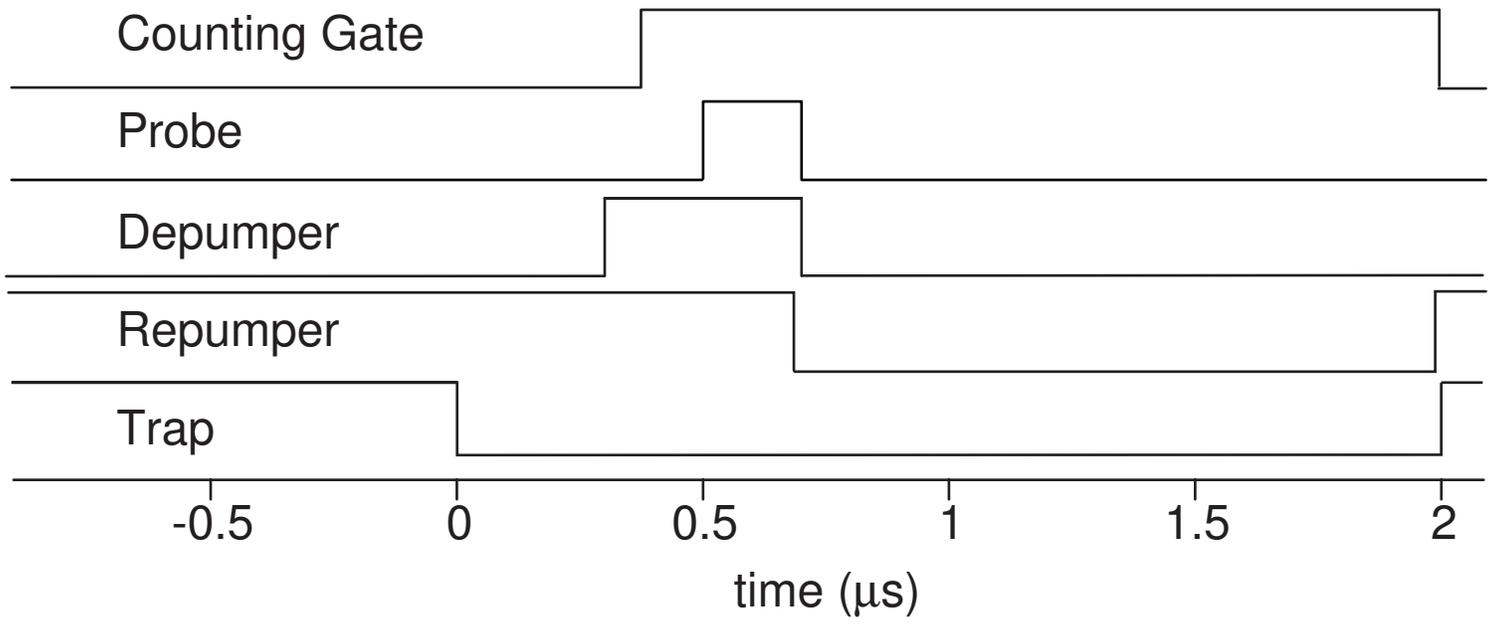}}}
\caption{Timing sequence for the excitation of atoms to the $7s$
level. High level is on, low level is off. gomezF4.EPS.
\label{Timing}}
\end{figure}

\noindent Fig. \ref{Timing}. Timing sequence for the excitation of
atoms to the $7s$ level. High level is on, low level is off.

\begin{figure}[h]\centerline{\scalebox{1}{\includegraphics{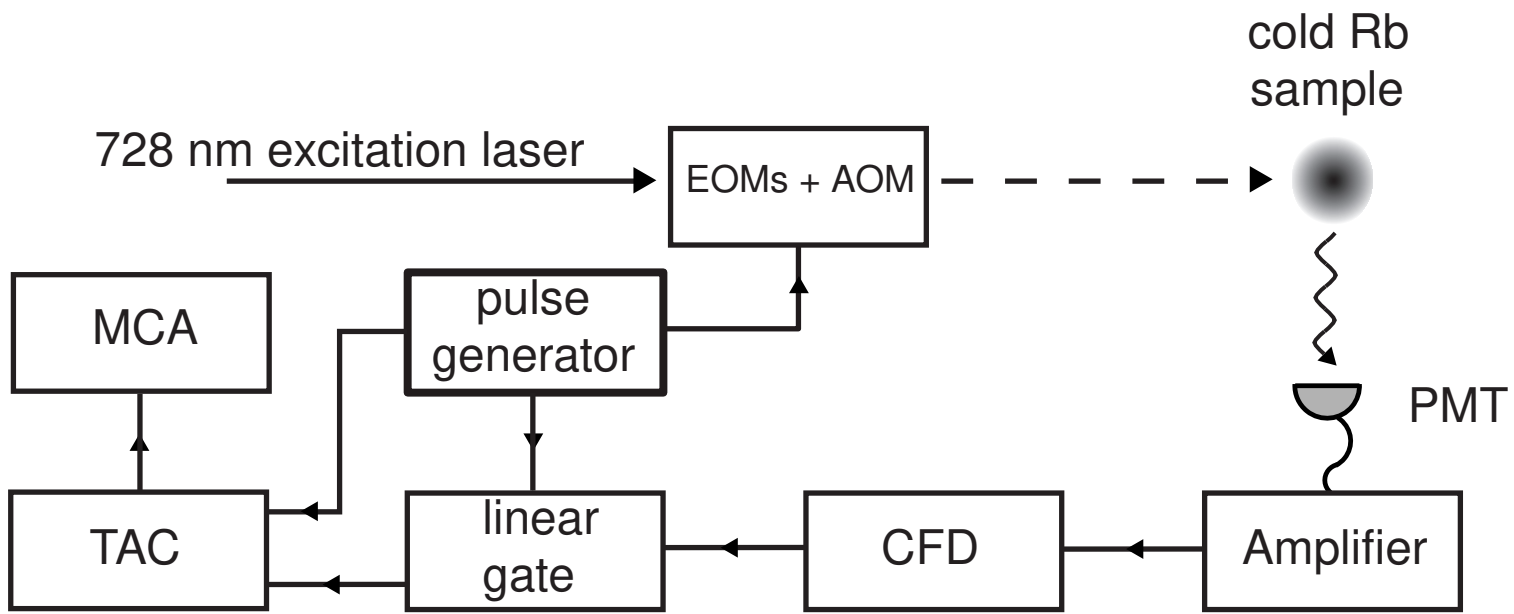}}}
\caption{Block diagram for the electronics used for the detection
of $7s$ or $6p$ photons. gomezF5.EPS. \label{Electronics}}
\end{figure}

\noindent Fig. \ref{Electronics}. Block diagram for the
electronics used for the detection of $7s$ or $6p$ photons.

\begin{figure}[h]\centerline{\scalebox{1}{\includegraphics{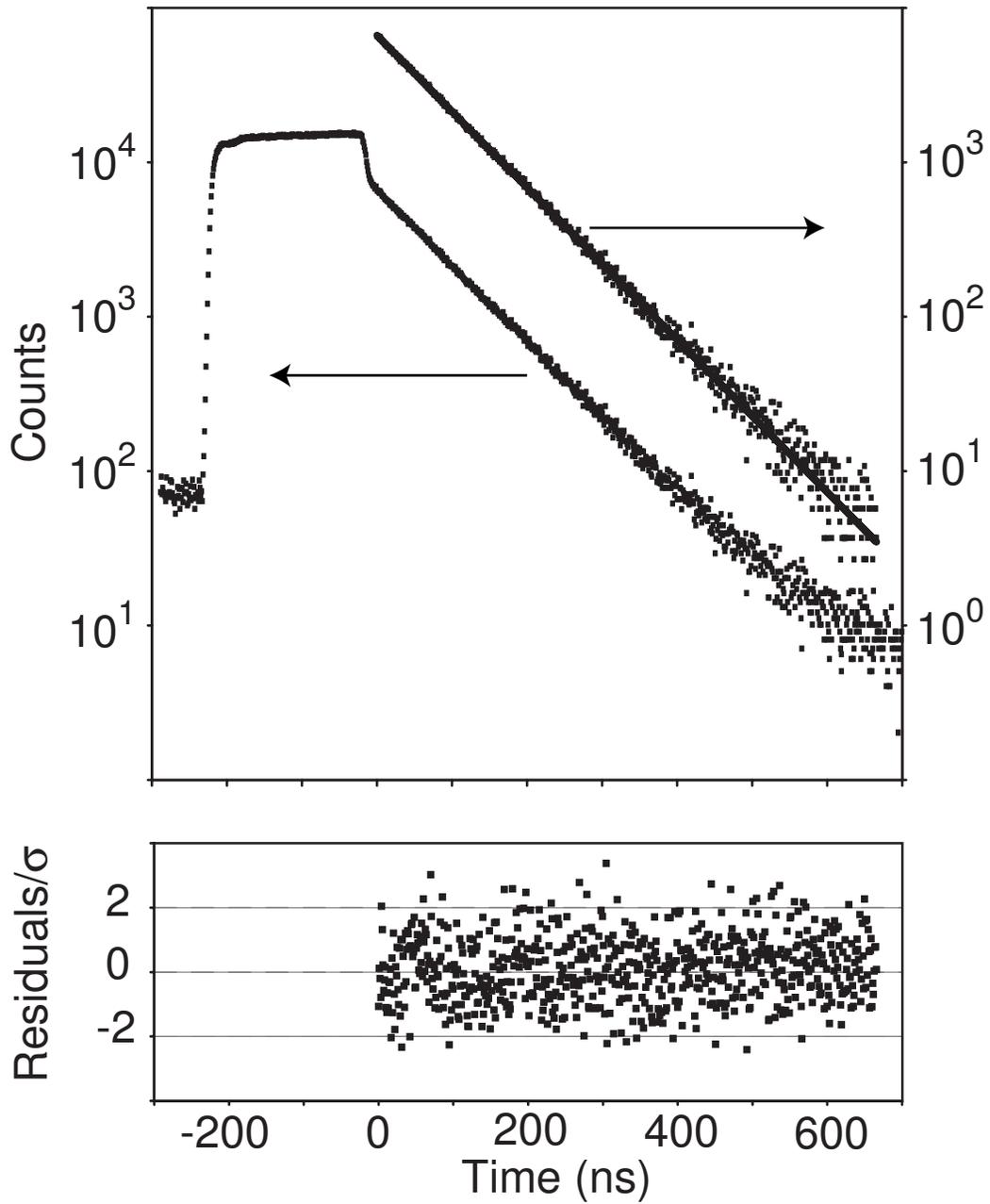}}}
\caption{Exponential decay of the $7s$ level. The upper plot
contains the raw data that shows the excitation turn on and turn
off as well as the exponential decay of the atoms (left scale). It
also shows the background subtracted signal together with the
exponential fit (right scale). The lower plot shows the normalized
residuals (assuming statistical noise). gomezF6.EPS.
\label{Decayrbres2}}
\end{figure}

\noindent Fig. \ref{Decayrbres2}. Exponential decay of the $7s$
level. The upper plot contains the raw data that shows the
excitation turn on and turn off as well as the exponential decay
of the atoms (left scale). It also shows the background subtracted
signal together with the exponential fit (right scale). The lower
plot shows the normalized residuals (assuming statistical noise).

\begin{figure}[h]\centerline{\scalebox{1}{\includegraphics{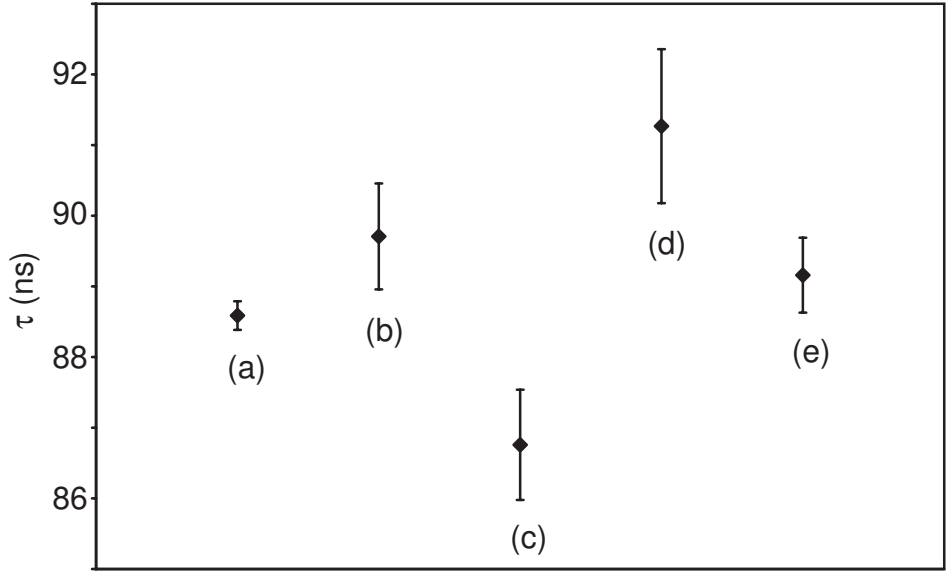}}}
\caption{Lifetime obtained when the trap is displaced by inserting
a piece of glass in the retro-reflection mirrors of the MOT while
the magnetic field environment remains unchanged. (a) no
displacement, (b) displacement using mirror 1, (c) mirror 2, (d)
mirror 3 and beams realigned, (e) no displacement and beams not
realigned. gomezF7.EPS. \label{Move}}
\end{figure}

\noindent Fig. \ref{Move}. Lifetime obtained when the trap is
displaced by inserting a piece of glass in the retro-reflection
mirrors of the MOT while the magnetic field environment remains
unchanged. (a) no displacement, (b) displacement using mirror 1,
(c) mirror 2, (d) mirror 3 and beams realigned, (e) no
displacement and beams not realigned.

\begin{figure}[h]\centerline{\scalebox{1}{\includegraphics{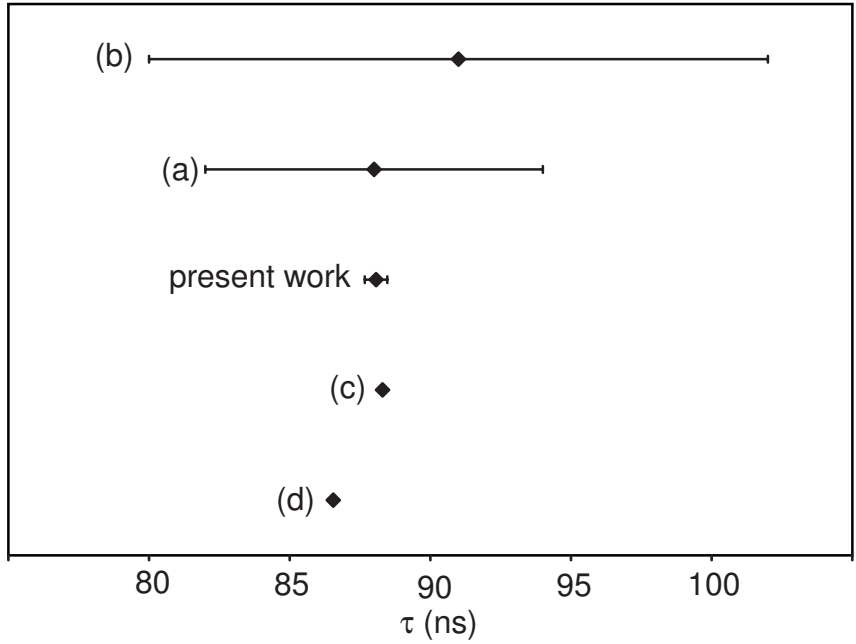}}}
\caption{Experimental result of the 7s lifetime in Rb, together
with previous experimental results: (a)Marek {\it et
al},\protect\cite{marek80} (b)Bulos {\it et
al},\protect\cite{bulos76} and theoretical predictions:
(c)Safronova {\it et al},\protect\cite{safronova03}
(d)Theodosiou.\protect\cite{theodosiou84} gomezF8.EPS.
\label{Theory}}
\end{figure}

\noindent Fig. \ref{Theory}. Experimental result of the 7s
lifetime in Rb, together with previous experimental results:
(a)Marek {\it et al},\protect\cite{marek80} (b)Bulos {\it et
al},\protect\cite{bulos76} and theoretical predictions:
(c)Safronova {\it et al},\protect\cite{safronova03}
(d)Theodosiou.\protect\cite{theodosiou84}

\begin{figure}[h]\centerline{\scalebox{1}{\includegraphics{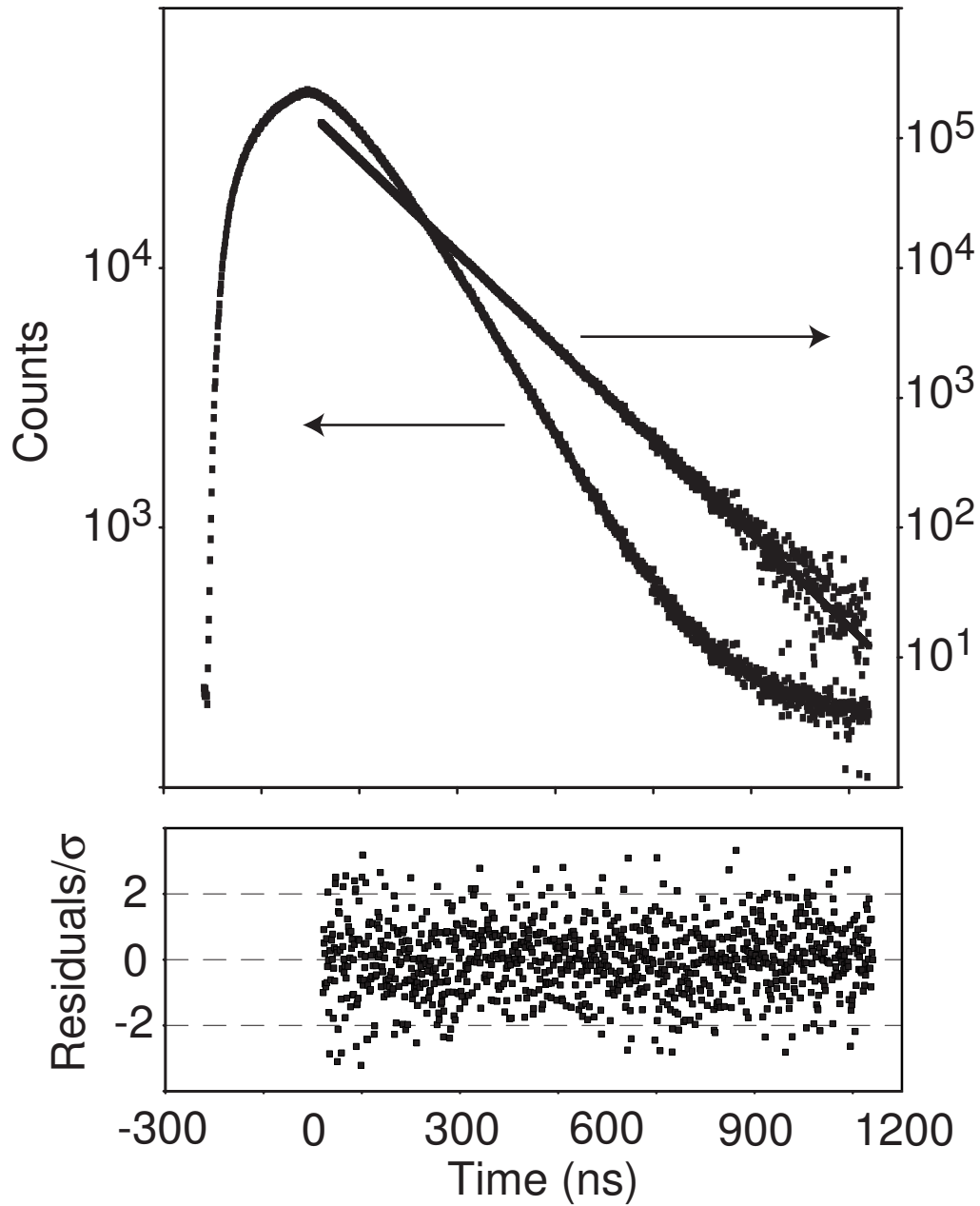}}}
\caption{Decay of the $6p$ manifold. The upper plot contains the
raw data (left scale) and the data minus the background minus the
exponential contribution from the $7s$ level (right scale). An
exponential fit to this last curve is also shown. The lower plot
contains the normalized residuals (assuming statistical noise).
gomezF9.EPS. \label{Decayresrb6p}}
\end{figure}

\noindent Fig. \ref{Decayresrb6p}. Decay of the $6p$ manifold. The
upper plot contains the raw data (left scale) and the data minus
the background minus the exponential contribution from the $7s$
level (right scale). An exponential fit to this last curve is also
shown. The lower plot contains the normalized residuals (assuming
statistical noise).

\begin{figure}[h]\centerline{\scalebox{1}{\includegraphics{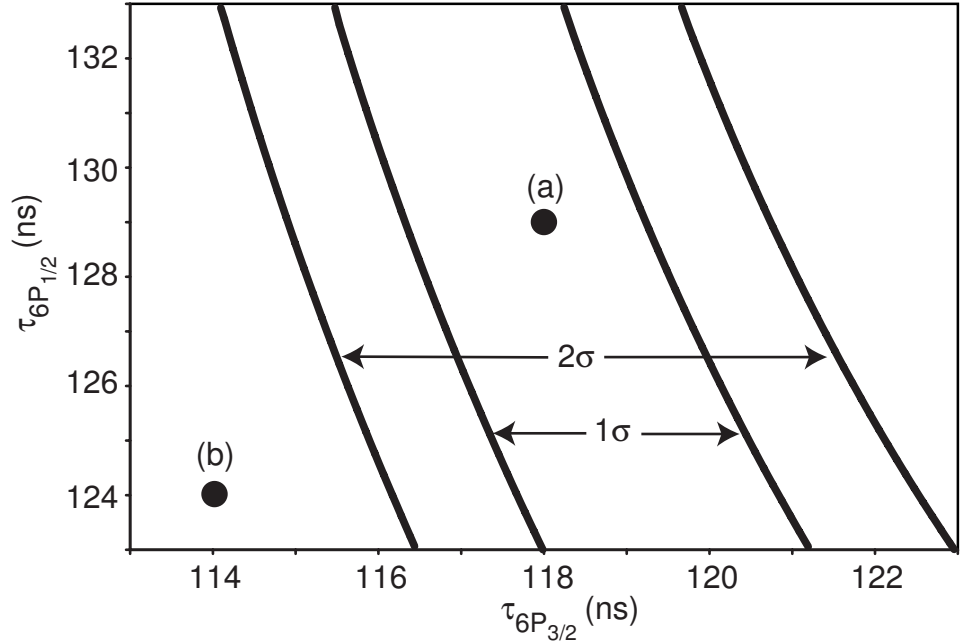}}}
\caption{Constraints on the lifetimes of the two $6p$ fine levels
in Rb using the model described in the text and the experimental
result. The solid lines define the limits of the 1$\sigma$ and
2$\sigma$ regions respectively. The circles are theoretical
predictions: (a)Safronova {\it et al},\protect\cite{safronova03}
(b)Theodosiou.\protect\cite{theodosiou84} gomezF10.EPS.
\label{6pprediction}}
\end{figure}

\noindent Fig. \ref{6pprediction}. Constraints on the lifetimes of
the two $6p$ fine levels in Rb using the model described in the
text and the experimental result. The solid lines define the
limits of the 1$\sigma$ and 2$\sigma$ regions respectively. The
circles are theoretical predictions: (a)Safronova {\it et
al},\protect\cite{safronova03}
(b)Theodosiou.\protect\cite{theodosiou84}

\begin{figure}[h]\centerline{\scalebox{1}{\includegraphics{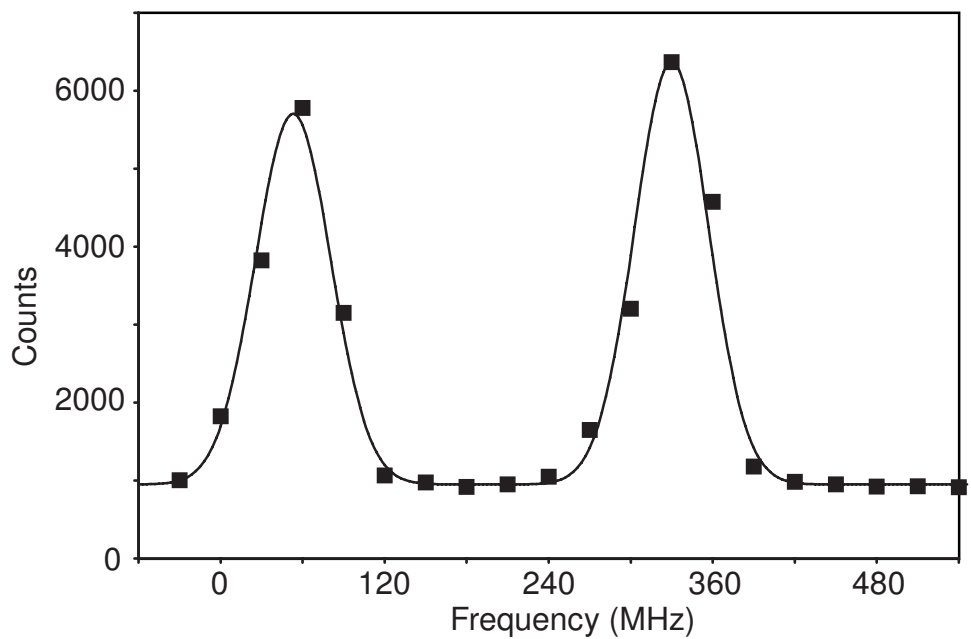}}}
\caption{Scan with wavemeter reading. The dots are the number of
photons per second and the solid line is a fit with two Gaussian
functions plus a background. The origin is arbitrarily defined to
be 13732.467 cm$^{-1}$ on the wavemeter. gomezF11.EPS.
\label{Wavemeter}}
\end{figure}

\noindent Fig. \ref{Wavemeter}. Scan with wavemeter reading. The
dots are the number of photons per second and the solid line is a
fit with two Gaussian functions plus a background. The origin is
arbitrarily defined to be 13732.467 cm$^{-1}$ on the wavemeter.

\begin{figure}[h]\centerline{\scalebox{1}{\includegraphics{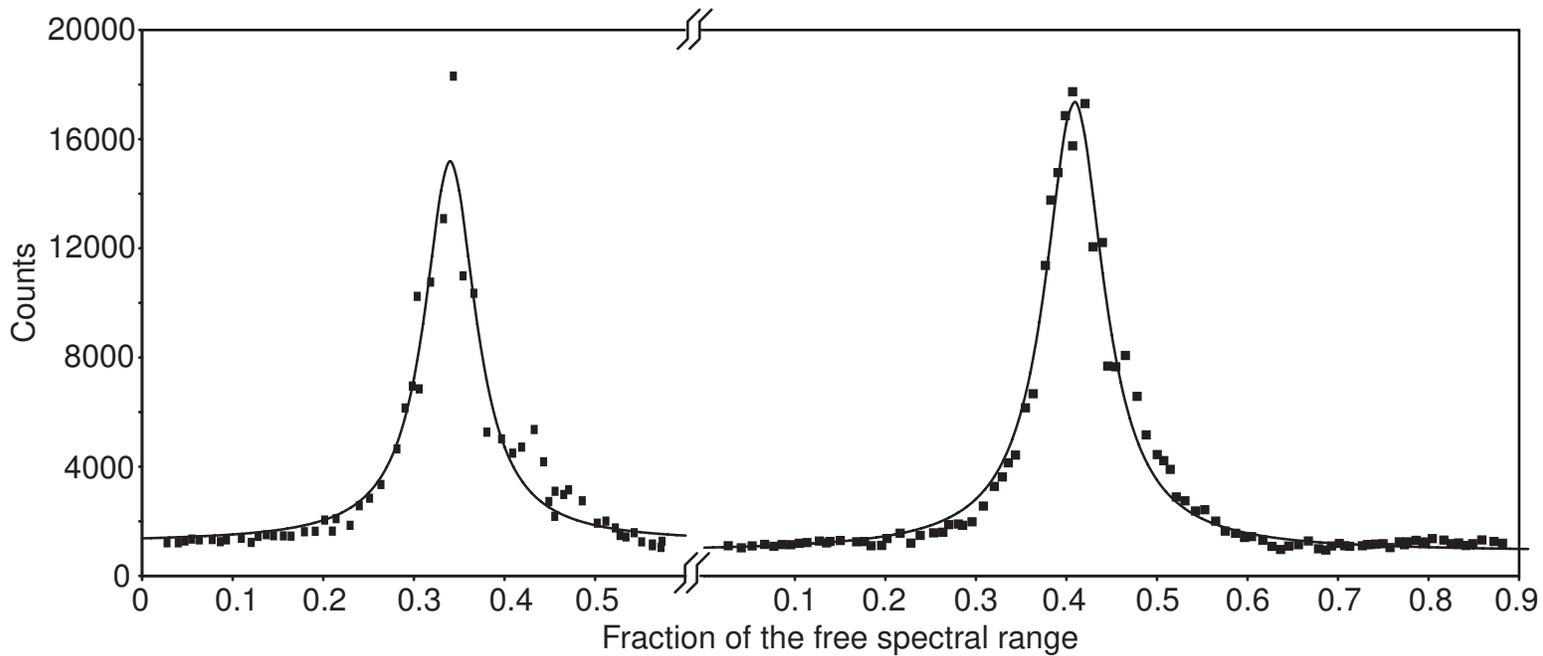}}}
\caption{Scan with the cavity reading. The horizontal axis is the
relative (or percent) position of the probe laser transmission
peak with respect to two fixed He-Ne transmission peaks in the
cavity. The dots are the number of photons per second and the
solid line is a fit with a Lorentz function plus a background. The
two peaks correspond to the two hyperfine levels and are separated
by one free spectral range. gomezF12.EPS. \label{Cavityscan}}
\end{figure}

\noindent Fig. \ref{Cavityscan}. Scan with the cavity reading. The
horizontal axis is the relative (or percent) position of the probe
laser transmission peak with respect to two fixed He-Ne
transmission peaks in the cavity. The dots are the number of
photons per second and the solid line is a fit with a Lorentz
function plus a background. The two peaks correspond to the two
hyperfine levels and are separated by one free spectral range.

\begin{figure}[h]\centerline{\scalebox{1}{\includegraphics{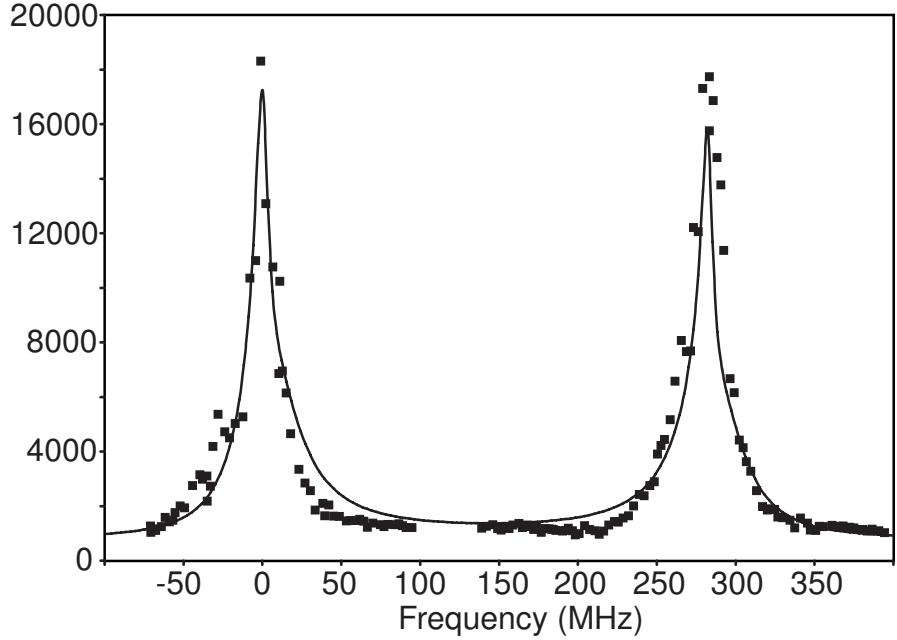}}}
\caption{Solution of the steady state optical Bloch equations and
its comparison with the data. The intensities and detunings of the
beams were adjusted to approximate the data and are consistent
with the experimental values. We also add a background and an
overall scale to the simulation. The probe intensity used is 27
mW/cm$^2$, the repumper intensity is 37 mW/cm$^2$ and the repumper
detuning is 3 MHz. gomezF13.EPS. \label{Simulation}}
\end{figure}

\noindent Fig. \ref{Simulation}. Solution of the steady state
optical Bloch equations and its comparison with the data. The
intensities and detunings of the beams were adjusted to
approximate the data and are consistent with the experimental
values. We also add a background and an overall scale to the
simulation. The probe intensity used is 27 mW/cm$^2$, the repumper
intensity is 37 mW/cm$^2$ and the repumper detuning is 3 MHz.

\begin{figure}[h]\centerline{\scalebox{1}{\includegraphics{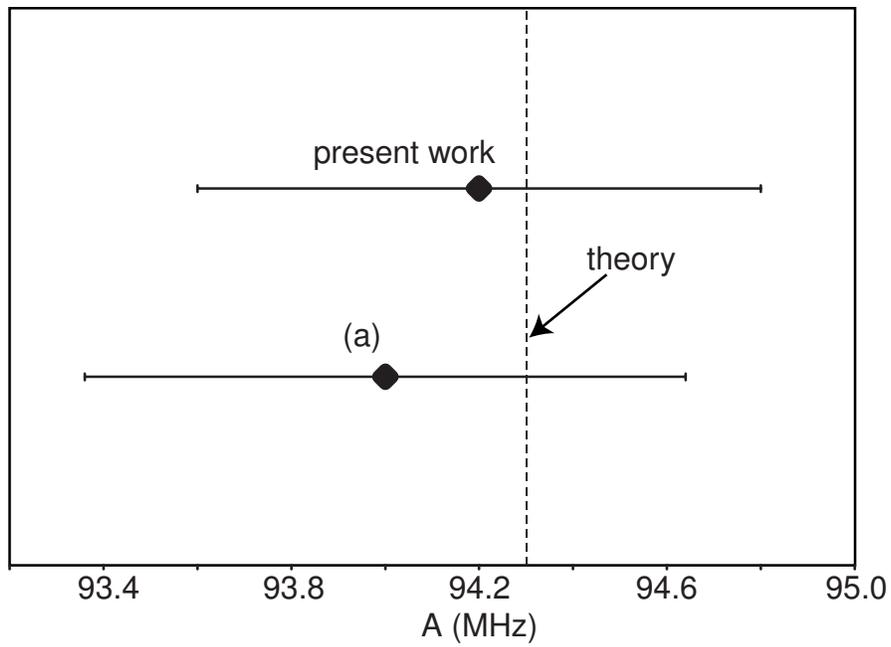}}}
\caption{Result for the hyperfine constant measurement and
comparison with: (a) previous experimental
result\protect\cite{gupta73} and theoretical prediction
\protect\cite{safronova99} (dotted line). gomezF14.EPS.
\label{Theoryhfs}}
\end{figure}

\noindent Fig. \ref{Theoryhfs}. Result for the hyperfine constant
measurement and comparison with: (a) previous experimental
result\protect\cite{gupta73} and theoretical prediction
\protect\cite{safronova99} (dotted line).

\newpage

\newpage

\newpage

\newpage

\newpage

\newpage

\newpage

\newpage

\newpage

\newpage

\newpage

\newpage

\newpage

\newpage

\end{document}